# THE ASTRONOMICAL SIGNIFICANCE OF 'NILURALLU', THE MEGALITHIC STONE ALIGNMENT AT MURARDODDI IN ANDHRA PRADESH, INDIA

**N. Kameswara Rao**
*Indian Institute of Astrophysics, Bangalore 560034, India.*
E-mail: nkrao@iiap.res.in

**Priya Thakur**
*P.G.Department of Studies and Research in History and Archaeology,
Tumkur University,Tumkur 572102, India.*
E-mail: Priya912@gmail.com

and

**Yogesh Mallinathpur**
*Deccan College Post Graduate Research Institute, Pune 411006, India*
yogesharchaeology@gmail.com

**Abstract:** The stone alignment 'Nilurallu' at Murardoddi is a megalithic monument containing standing stones of 12 to 16 feet high that are arranged somewhat in a squarish pattern. This is one of the stone alignments listed by Allchin (1956) as a non-sepulchal array that might have some astronomical connotations. This impressive stone alignment seems to be similar to that at Vibhuthihalli, that was studied earlier, but constructed with much larger stones. The observations conducted by us show that the rows of stones are aligned to the directions of sunrise (and sunset) on calendrically-important events, like equinoxes and solstices. In contrast to Vibhuthihalli, the shadows of stones provide a means of measuring shorter intervals of time.

## 1 INTRODUCTION

On the day of the spring equinox (e.g. 21 March 2010) people from Mudumala and Murardoddi in India visit 'Nilurallu' (which means 'standing stones' in Telugu), a stone alignment located between these two villages, offer their prayers and celebrate the occasion by having a feast. Nilurallu is a megalithic alignment of standing and fallen stones that are 12-16 feet (3.7-4.9 meters) high (Figure 1) in the Mahbubnagar district of Andhra Pradesh. The celebration (Figure 2) in a way demonstrates a continuing tradition commemorating an astronomical event and its connection with this alignment. This is one of the megalithic alignments listed by Allchin (1956) as non-sepulchral stone arrays with possible orientations towards cardinal points, similar to Vibhuthihalli. We have surveyed about 17 of the sites listed by Allchin (Table 1) out of which only the Vibhuthihalli and Murardoddi alignments are relatively undisturbed. The remainder have either disappeared completely, or have very few stones left. General properties of the sites have been described by Allchin. The plan consists of stones arranged in parallel rows with equal spacing. The stone arrangements are either square-like, a checker board, or a square with a diagonal arrangement consisting of one more stone in the centre of a mini-square formed from a set of four stones. The effect is to stress the diagonals.

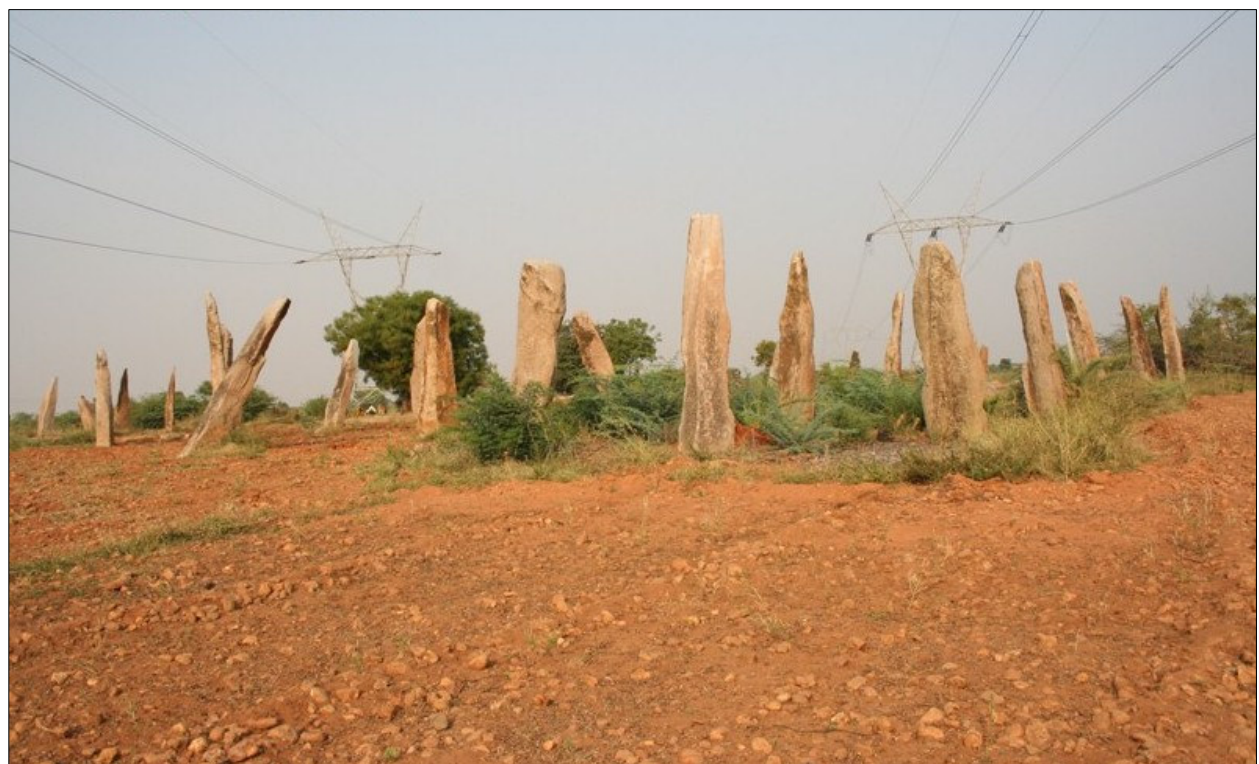

Figure 1: The overall view of the Nilurallu alignment as seen from the western side.

Table 1: Status of the megalithic sites visited.

| Sl No. | Site and Location | Type of Megalith | Comments |
|---|---|---|---|
| 1. | Vibhuthihalli, 16°.665 N, 76°.858 E; Shahput Taluk, Yadgir, Karnataka | Stone Alignment | Preserved |
| 2. | Bheemarayanagudi, 16°.727 N, 76°.798 E; Shahput Taluk, Yadgir, Karnataka | Stone Alignment | Disturbed |
| 3. | Ijeri, Shahput Taluk, Yadgir, Karnataka | Stone Alignment | Does not exist |
| 4. | Rajan Kollur, 16°.39 N, 76°.455 E Shorapur Taluk, Yadgir, Karnataka | Stone Alignment Dolmens | Does not exist Exists |
| 5. | Hanamsagar, 15°.883 N, 76°.072 E; Shorapur Taluk, Yadgir, Karnataka | Stone Alignment | Disturbed |
| 6. | Managodanahalli, Devanahalli Taluk, Bangalore, Karnataka | Menhirs, Cists | Does not exist |
| 7. | Koiera, Shorapur Taluk, Yadgir, Karnataka | Menhirs, Cists Dolmens | Does not exist Exists |
| 8. | Mudumala 16°.379 N, 77°.41 E; MakhtalTaluk, Mahabubnagar, Andhra Pradesh | Stone Alignment | Exists to some extent |
| 9. | Murardoddi, 16°.378 N, 77°.406 E; Makhtal Taluk, Mahabubnagar, Andhra Pradesh | Stone Alignment | Exists |
| 10. | Panjanur (Pundununnur), 16°.386 N; Makhtal Taluk, Mahabubnagar, Andhra Pradesh | Habitation | Disturbed |
| 11. | Gudabellur, 16°.42 N, 77°.383 E; Makhatal Taluk, Mahabubnagar, Andhra Pradesh | Stone Alignment | Does not exist |
| 12. | Kotakunda-Koilkunda, Makhatal Taluk, Mahabubnagar, Andhra Pradesh | Stone Alignment | Does not exist |
| 13. | Madhawavaram, Makhatal Taluk, Mahabubnagar, Andhra Pradesh | Stone Alignment | Does not exist |
| 14. | Gopalpur, Makhatal Taluk, Mahabubnagar, Andhra Pradesh | Stone Alignment | Does not exist |
| 15. | Devakadra, 16°.616 N, 77°.833 E; Makhatal Taluk, Mahabubnagar, Andhra Pradesh | Stone Alignment | Does not exist |
| 16. | Kundanpur-Sanganunpalli, Makhatal Taluk, Mahabubnagar, Andhra Pradesh | Stone Alignment | Does not exist |
| 17. | Jamshed I - IV, 16 Raichur, Karnataka | Stone Alignments | Do not exist |
| 18. | Krishna Bridge, Raichur, Karnataka | Stone Alignment | Does not exist |

In an earlier paper (Kameswara Rao and Thakur, 2010; hereafter Paper I) we showed that the Vibhuthihalli stone alignment, in all likelihood, was used as a calendarical device by megalithic people from the Karnataka region. In the present paper we explore the possible astronomical significance of the Nilurallu alignment by monitoring sunrises and sunsets during calendrically-important occasions, such as the equinoxes and the solstices. Most of the alignments listed by Allchin have stones of 4-6 feet (1.2-1.8 meters) and 2-4 feet (0.6-1.2 meters) in diameter. Thus, the Nilurallu alignment is special as it consists of huge stones, two to three times the usual size. It might have had a special purpose as well.

## 2 THE SITE

The Nilurallu alignment is located at latitude 16° 22' 44" N and longitude 77° 24' 40" E, southeast of Murardoddi village and southwest of Mudumala village. The Krishna River is about a 0.5 km away to the south (Figure 3). The alignment is in an area containing artifacts of different cultures, ranging in age from the Middle Paleolithic and Mesolithic to the Megalithic Period. In the course of our field work we found stone circles and avenues of smaller (i.e. normal sized) stones in the area to the east of Nilurallu (also see Krishna Sastry, 1983). Areas to the west of Nilurallu have some historical hero tablets, sculptures, iron slags, bruisings and even small rock pits used for sharpening stone tools (ibid.). To the south there are arrays of 0.5 to 3 feet (0.15 to 0.9 meter) high stones (ibid.). Meanwhile, Middle Palaeolithic tools (choppers, and different types of scrappers, borers, and flake tools) have been recovered from the site of the Nilurallu alignment (ibid.). The site and its surroundings are filled with pebbles. The Nilurallu site occupies a prominent place on slightly elevated ground that dominates the surrounding terrain.

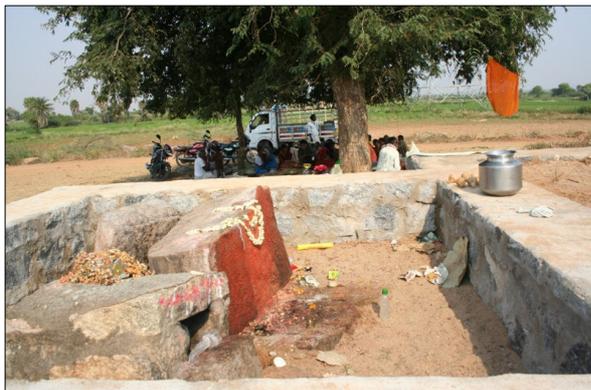

Figure 2: A celebration at the Nilurallu stone alignment site on equinox day.

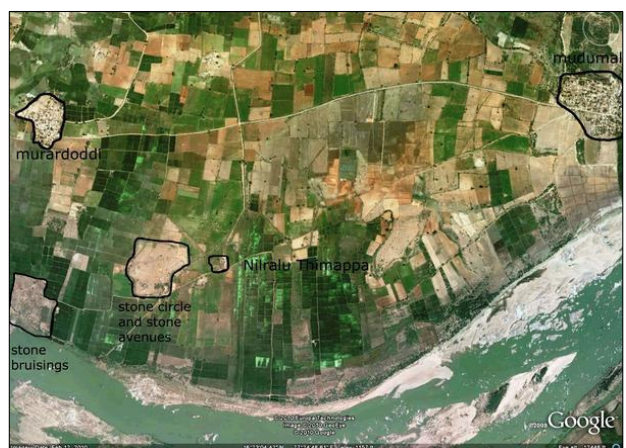

Figure 3: Google map of the area with the stone alignment. The Krishna River is about half a kilometer away to the south.

The earliest account of the site is by Krishna Murthy (1941: 86), who describes it as follows:

> … there is an almost square area studied with rough-hewn stone pillars. These pillars are arranged in parallel rows in a north-south direction. The pillars are 14′-16′ long and 6′-11′ in girth. There are 31 pillars still standing and many have fallen down. The square measures about 200′ × 200′ with apparently 6 pillars in each row.

Krishna Murthy (1941) also provides a photograph (which, unfortunately, we could not obtain). He adds: "… the pillars are locally known as 'Nilu Ralloo' meaning standing stones." (ibid.). Krishna Murthy (1941) also provides a local legend for the origin of Nilurallu alignment. Apparently a disappointed old beggar woman, who was deceived by the local farmers while harvesting grain, cursed them and they became stones. The standing pillars represent the men who were working and the fallen ones are those people who were lying down. The large group of short stunted pillars to the southwest of this alignment are supposed to be petrified cattle, and the sand around the stones was supposedly the grain they were harvesting. Krishna Murthy (1941) also seems to have picked up a piece of a stone axe at the site. Allchin (1956) quotes Krishna Murthy's description in his report. To our knowledge, these are the only reports about this site that have been published.

However, recently, two abstracts of conference papers by K. Pulla Rao (2007; 2009) appeared that provide some description of his observations of the site. According to him there are "… more than 800 menhirs arranged in different formations and rows. The rows are oriented in different directions." (Pulla Rao, 2009). He also mentions that observations on

> … summer and winter solstice reveals that one particular row aligns with the Sun in the morning and another row in the evening ... [and] The central area of the complex has a concentration of about 80 tall (up to 14 feet) menhirs which are arranged in rows forming alignments and avenues. The rows are oriented in different directions. (ibid.).

These bigger menhirs are the ones Krishna Murthy (1941) described earlier, mentioning that they were arranged in parallel rows but were not oriented in specific directions. In one of the abstracts, Pulla Rao (2007) mentions that "The central area with the bigger menhirs also has a formation of stones arranged in concentric circles with standing menhirs interspersed with horizontal blocks. It has been observed that two of the taller menhirs of the circle align with the Sun in both morning and evening." However, he does not mention on which days this alignment occurs. Presently no such structure of "concentric circles" of bigger menhirs exists, nor does Krishna Murthy (1941) mention any such formation.

Pulla Rao (2007) also found a stone in the southwestern part of the complex which has 30 cup marks on its surface. He claims that these cup marks depict the constellation Ursa Major. We have located this rock (Figure 4) in the southwest periphery. The rock surface contains about fifty cup marks, not just seven. A comparison with sky charts (e.g. in Norton's Star Atlas—see Ridpath, 2004) of the naked eye stars shows that any resemblance to the Great Bear constellation is imaginary. There is no indication that these marks are even depictions of stars. They could even represent the layout of the stone arrays! At this same site we also found another stone with several cup marks on it, as well as a stone with a solitary cup mark. Stones with cup marks are found at other megalithic sites (see Peddayya, 1976).

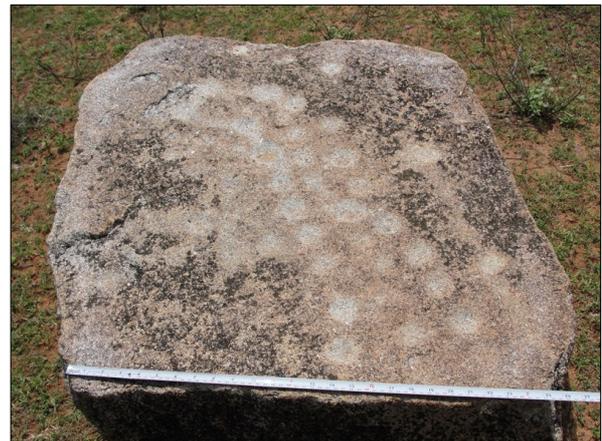

Figure 4: It has been claimed that this bruising (the cup marks) depict the constellation of Great Bear (Ursa Major).

## 2.1 Recent Observations

The Nilurallu site is on private land and the owners plant crops when they desire, so astronomical observations have to be conducted when the fields are empty. Figure 5 illustrates the comparative size of some of the stones relative to people standing next to them. As of July 2009, there were 29 standing stones and 46 fallen ones, whereas a little over a year earlier—in April 2008—there were 31 to 32 standing stones. Thus, two or three standing stones disappeared during this interval. Some of the fallen stones have been heaped on the eastern side of the site between standing stones s27 and s28. All of the standing stones present in July 2009 and most of the fallen ones that are lying in open areas (and not heaped up) are plotted in Figures 6 and 7. A scaled map was generated from our measurements of the spacing between the different stones. This map is also consistent with the appearance of an enlarged Google Earth map of the region, except for some fallen stones in the eastern part of the site that were not measured by us.

The height of the standing stones varied from 3.0 meters to 4.7 meters. The stones were mostly granite, and they taper towards the top. They are of different shapes. In most cases their thickness is smaller than their width, thus showing a slab-like appearance (e.g.

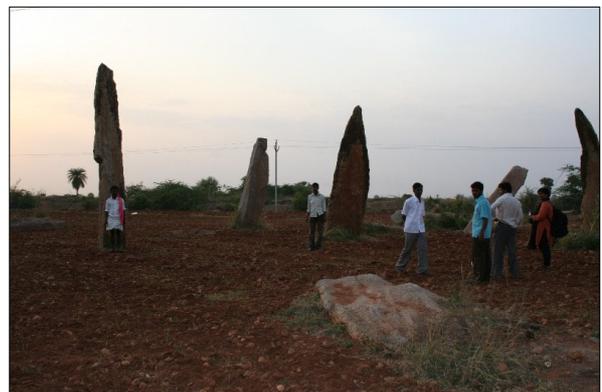

Figure 5: Some of the stones are shown with people around them to illustrate the height and size of the stones.

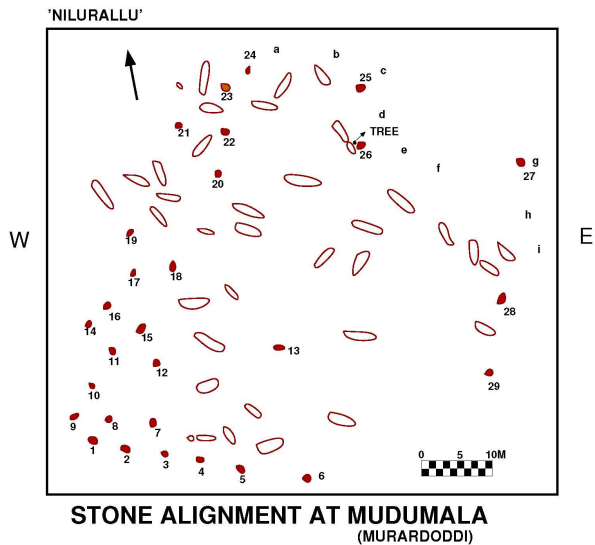

Figure 6: Scaled map of the Nilurallu site layout as measured by us. The filled spots that are numbered represent the standing stones and the open contours refer to the fallen stones. The numbers were arbitrarily given by us for the purposes of this study.

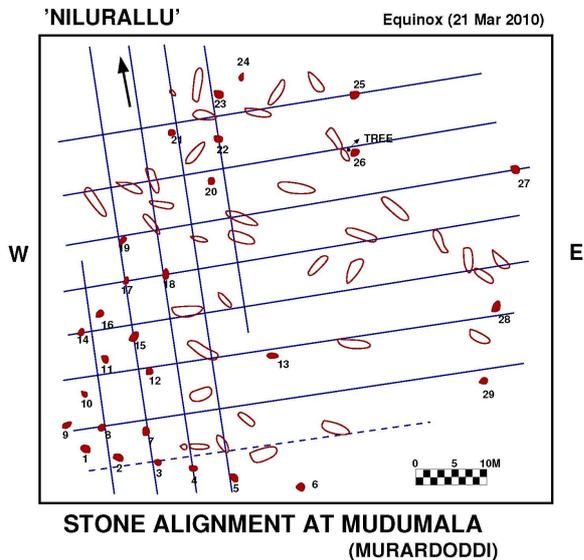

Figure 7: A map of the Nilurallu layout with the directions of sunrise and sunset at the time of the equinoxes shown by the near-horizontal lines. The near-vertical lines indicate the north-south direction.

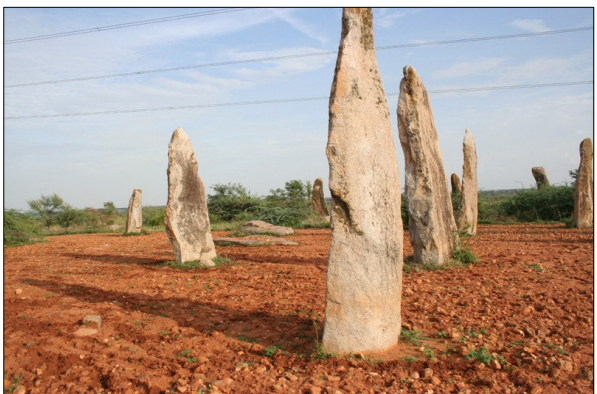

Figure 8: A row of stones in a north-south direction. The stones numbered 19, 17, 15 and 12 (centre of the figure) can be seen to lie in a straight line. See Figures 6 and 7 for the numbering. This photograph was taken on 21 June 2010 at 5:17 pm (IST).

s22, s18, s15, s11 etc.), although some are like cylinders (e.g. s25). The stone numbers referred to are shown in Figures 6 and 7 and are internal to this study. We have marked these numbers on the stones too, for future reference during this study. Several of the stones are tilted from the vertical (the figures show the base positions) and give an impression that the tilt is inwards if looked at from afar (e.g. see Figure 1).

### 2.2 The Archaeological Setting

Because the site is in a slightly elevated area, the horizon in all directions is quite clear, and sunrises and sunsets are readily visible—except when there are crops on the site. The present extent of the site seems to be the same as that mentioned by Krishna Murthy (1941), i.e. about 200 ft × 200 ft. The layout as mapped gives the impression of a diagonal alignment. The separation of rows of stones seems to be uniform more or less (but defining them is not always easy and depends mostly on the presently-standing stones). The average separation of any two stones in a row (centre to centre) is 5.8 ± 0.9 meters.

#### 2.2.1 North-South Direction

We determined the north-south line, as in Vibhuthihalli, by using a stick as a gnomon to measure the direction of the shadow of the stick. The meridian direction was established by marking on the ground the direction of the shortest shadow (i.e. when the Sun was on the meridian).

As far as possible, we tried to adopt methods which were simple and could have been accomplished with tools that were available in prehistoric times. Figure 7 shows the direction and the rows of stones which are parallel to north-south. A specific example, illustrated in Figure 8, is the row with stones s19, s17, s15, s12, s7 and s3 that lie in a north-south direction.

### 3 CALENDRICAL EVENTS

One of the important aspects regarding the astronomical relevance of the monument is to observe whether preferred alignments exists for calendrically-important events like sunrises and sunsets on equinox and solstice days. Were the rows of stones aligned in these preferred directions? We tried to monitor sunrises and sunsets on a few days around the dates of the equinoxes and solstices, weather permitting, and visual observations were made along the rows of standing stones pointing towards the directions of sunrise and sunset. Since the tops of the stones are tilted in random directions (in some cases), the positions of the bases of the stones, which were less disturbed, were used to define the directions.

### 3.1 Equinox Sunrises and Sunsets

Observations were made at the times of the September 2009 and March 2010 equinoxes. However, the September observations were mostly hindered by clouds. The equinox occurred on 20 March 2010 at 23:30 (IST), and our observations extended from 19 to 22 March. The sunrises and sunsets along the rows are marked in Figure 7. Examples of the direction of sunrise along various rows are illustrated in Figures 9, 10 and 11. Since the view directly along a row would be blocked by the stone in front, views from slightly to the right or the left of the stone in front are illustrated. The accuracy of the stated direction for each row is ~1° (i.e. two solar diameters). Since the rows point to the equinoxial sunrise (in the east) and sunset (in the west) they are parallel to each other and are shown as horizontal lines in Figure 7. As an example the row of stones, s9, s8, s7 and s29 points to the equinox sunrise (s29 is hidden in the Figure 9); as does the row with s12, s13 and s28 and the row with s17, s18. The rows as drawn in Figure 7 from the directions of sunrise and sunset provide roughly equal spacing between them and are perpendicular to the north-south rows. Equinoxial sunsets along the rows of stones are illustrated in Figures 12, 13 and 14. The calculated azimuth and altitude of the Sun are also given in the Figure captions.

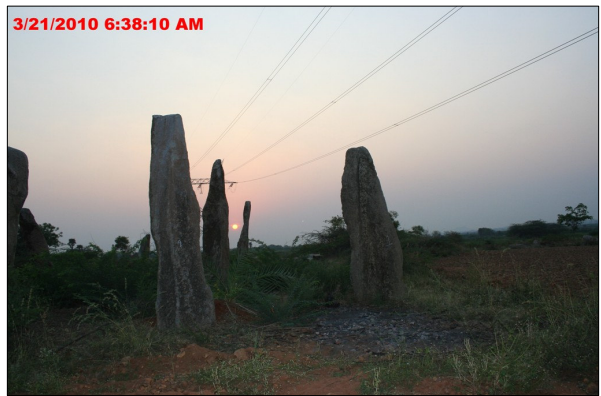

Figure 9: Equinoxial sunrise as seen over the stone row containing stones numbered s9, s8, s7 and s29. The Sun's calculated azimuth is 89.39° and altitude is 2.52°.

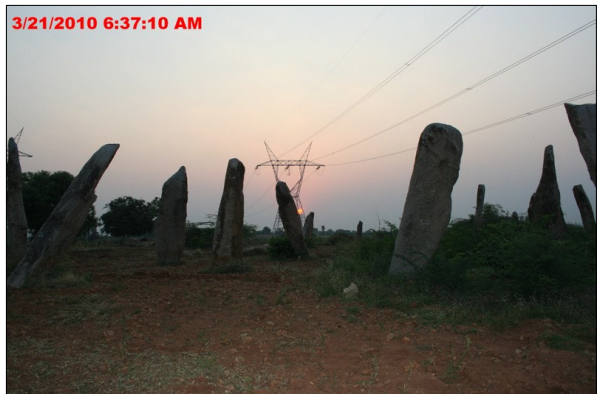

Figure 10: Equinoxial sunrise as seen over the stone row containing stones numbered s12, s13 and s28. The Sun's calculated azimuth is 89.5° and altitude is 2.32°.

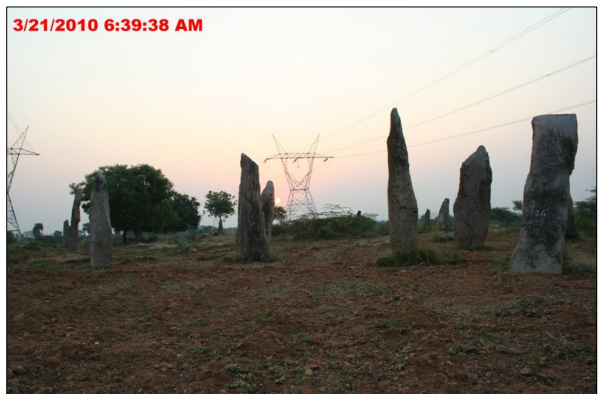

Figure 11: Equinoxial sunrise as seen over the stone row containing stones numbered s17 and s18. The Sun's calculated azimuth is 89.30° and altitude is 2.8°.

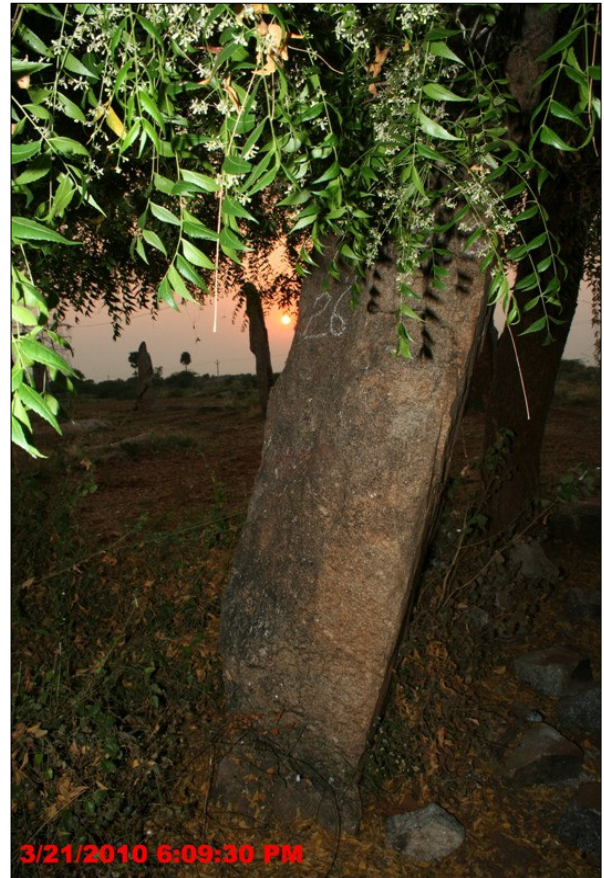

Figure 12: Equinoxial sunset over stones s26 and s20. The Sun's calculated azimuth is 270.8° and altitude is 4.4°.

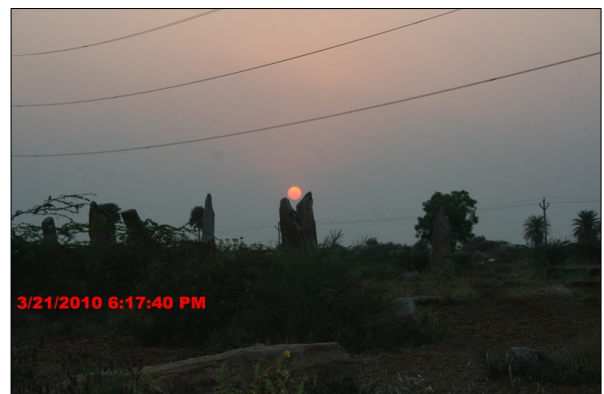

Figure 13: Equinoxial sunset over stones s18 and s17. The Sun's calculated azimuth is 270.4° and altitude is 2.5°.

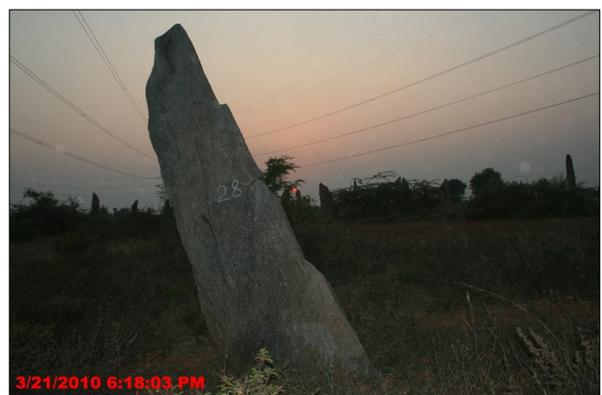

Figure 14: Equinoxial sunset over stones s28, s13 and s12. The Sun's calculated azimuth is 270.36° and altitude is 2.36°.

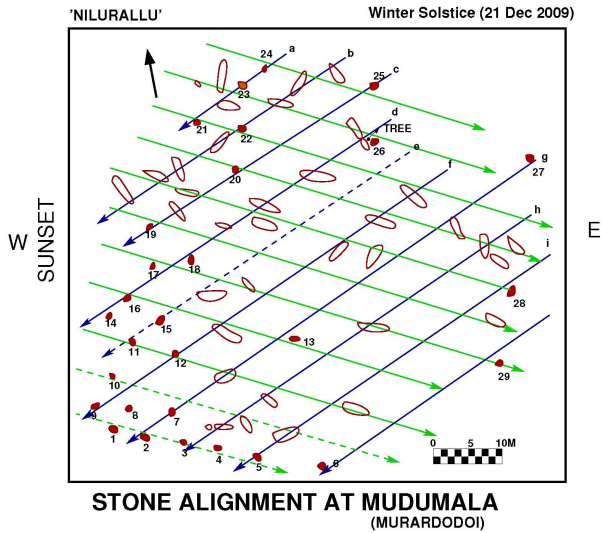

Figure 15: Sunrise (green lines) and sunset (blue lines) directions during the winter solstice are shown on the Nilurallu site layout. The diagonal stone rows showing the direction of sunset are denoted by the letters a, b, c, etc.

### 3.2 Solstice Sunrises and Sunsets

The solstice observations were obtained in December 2009 and June 2010. The summer solstice sunrise and sunset observations were affected by clouds. Winter solstice occurred on 21 December 2009, and we obtained observations from 19 to 22 December. The sunrise and sunset directions along the stones are marked in Figure 15. The stones seem to be well laid out in these directions and show a parallel outlay as expected. The images of sunsets along various diagonal rows are illustrated in Figures 16 to 19. Figure 17 shows the row with stones s25, s20 and, and s19 pointing to the sunset direction. Similarly, the sunrise direction is illustrated by s17 and s29 in Figure 18. There seem to be about ten diagonal and parallel rows present in the layout, as shown in Figure 15 suggesting a separation of about 20' between the rows (i.e. 6.0 meters). Even the parallel rows aligned with equinoxial sunrise and sunset might number the same, with similar spacing, as seen in Figure 7.

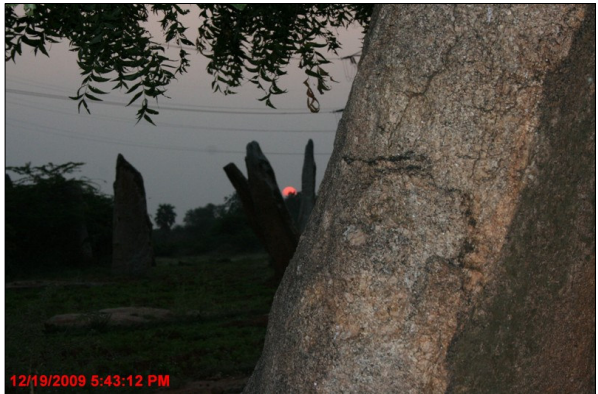

Figure 16: Winter solstice sunset over the row containing stones s26, s18, s16 and s14. The Sun's calculated azimuth is 245.16° and altitude is 1.1°.

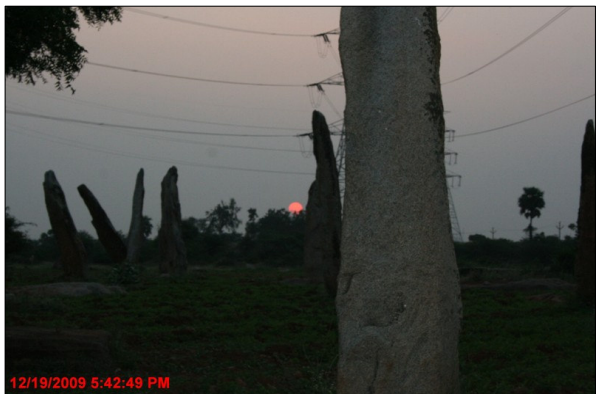

Figure 17: Winter solstice sunset over row c, containing stones s25, s20, and s19. The Sun's calculated azimuth is 245.14° and altitude is 1.18°.

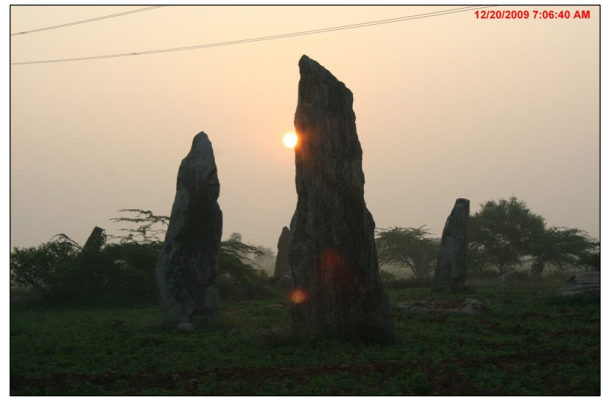

Figure 18: Winter solstice sunrise over stones s17 and s29. The Sun's calculated azimuth is 244.06° and altitude is 4.24°.

### 3.3 Shadows

Another major aspect of the Nilurallu alignment, that is not present at Vibhuthihalli, is the shadows of the stones. In particular, the shadows cast (by sunlight) on each other by high and broad neighbouring stones form a pattern that is so distinct and prominent that it enables the time during the day, as well as the day (like latter-day sundials), to be determined. Some examples are shown in Figures 20 to 23.

The stones are of sufficient height to cast prominent shadows not only on the neighbouring stones but also on the dry reddish soil underneath. The length and direction of the shadows change with time and day. At mid-day, when the Sun is on the meridian, a north-south row of stones would cast shadows which would be in a line (e.g. the stones s19, s17, s15, s12, etc.). On the winter solstice day the shadows at mid-day would be about half of the separation of stones (so a 12-foot high stone would have a shadow of about 10 feet in length).

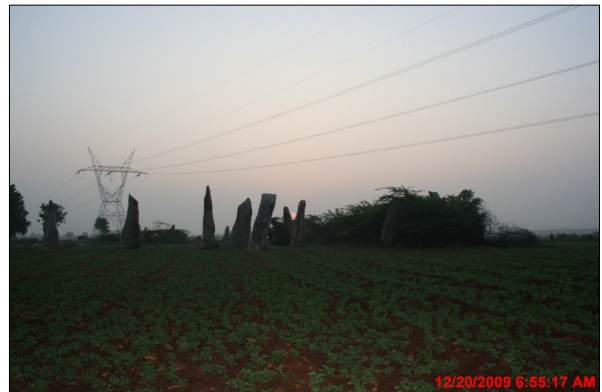

Figure 19: Winter solstice sunrise over stones s11 and s12. The Sun's calculated azimuth is 244.92° and altitude is 1.8°.

Figure 21 illustrates the change in shadow direction in the evening from summer solstice to equinox to winter solstice. At summer solstice the stone s11 casts its shadow such that it falls on s12 (Figure 21a). At the equinox, s11's shadow falls midway between stones s15 and s12 (see Figure 21b). At winter solstice the shadow of s11 falls on the edge of s15. The altitude of the Sun on these three occasions is in the range of 21-28°, and does not differ much from one date to another.

Thus shadow length and direction could provide the time of the day. Distinct markers could be made when a stone cast its shadow and fully covered a nearby stone, another marker when it reached its base, etc., to reckon the time.

## 4 THE ANATOMY OF THE NILURALLU STONE ALIGNMENT

It is very clear from the earlier discussion that the stone rows are distinctly aligned to sunrises and sunsets during both the equinoxes and the solstices. Several issues regarding the squarish plan—a general feature of all of the stone alignments in Karnataka, Andhra Pradesh region, listed by Allchin (1956)—have been discussed in Paper I (Kameswara Rao and Thakur, 2010), but they do apply to Nilurallu. Although the present site lacks the sharp-edged demarcations as seen at Vibhuthihalli, it does look to be a square.

### 4.1 The Plan

As was discussed in Paper I, the squarish plan is well suited to point to the directions of the solstices at Nilurallu. At the latitude range of 16-17° the azimuthal travel of the Sun (plus the size of the Sun's disk) on the horizon was slightly over 50° in 1500 BC (i.e. ~25° on each side of the equinox sunrise—or sunset—direction). This angle is the same as the angle measured from a centre of a side to the opposite corners of a square within the error of two solar diameters. The Nilurallu alignment is also consistent with this picture. The diagonal lines are drawn parallel to the solstice directions. In this picture, a preferred position would be the centre of either eastern or western side. But as the rows are parallel to each other the view of sunrise and sunset could be monitored from any row.

Why would one need to have so many stone rows to mark the major calendrical events? As discussed in Paper I, the Sun's motion on the horizon is not uniform. It accelerates as it approaches the equinoxes and slows down as it nears the solstices. To monitor this motion and to be able to predict the day (or how many days before or after the equinox or the solstice) more stone markers were required. Smaller increments of motion per day would need to be measured near the solstices and larger ones near the equinoxes.

### 4.2 Comparison of the Nilurallu and Vibhuthihalli Stone Alignments

Obviously the Nilurallu site has much bigger stones, about three times higher than those used at Vibhuthihalli. The extent of the site is also less by 3.5 times (720 feet, or 219.5 meters for Vibhuthihalli as compared to 200 feet or 61 meters for Nilurallu). The spacing between the stones at Nilurallu is half that at Vibhuthihalli (about 20 feet or 6.1 meters as compared to 38 ± 3 feet or 11.6 meters at Vibhuthihalli). The Nil-urallu alignment is more compact but is much more impressive because of the use of tall and massive stones. Although the main reasons for using such large stones is not known, one of the primary purposes could be to utilize the shadows of the stones as markers of time. The reason for reducing the spacing between the stones and increasing their height might be to enable the shadows to cover the adjacent stones. Many of the stones chosen are broad (more like slabs), thereby providing a flat surface upon which to see the shadow of the stone in front. Perhaps tapering of the stones towards their tops may have kept the shadows sharp and pointed so as to serve as markers. Another advantage of using tall stones is that they could be used as screens to directly view the Sun (at average human height). Even if shadows were not present one could still estimate the position of the Sun as it came out from behind the screen.

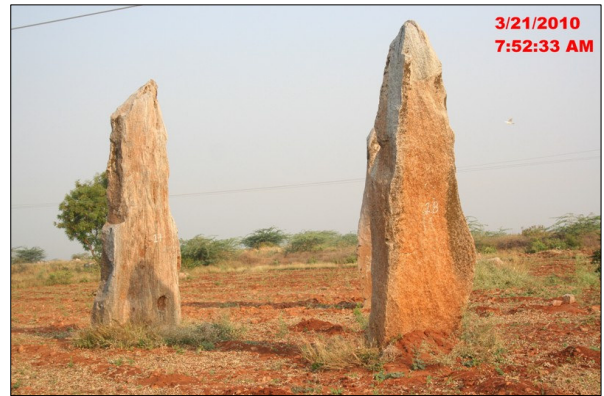

Figure 20: Shadows of stones on each other at the time of the equinox. The shadow of s18 on s17 is shown. The Sun's calculated azimuth is 83.88° and altitude is 20.41°.

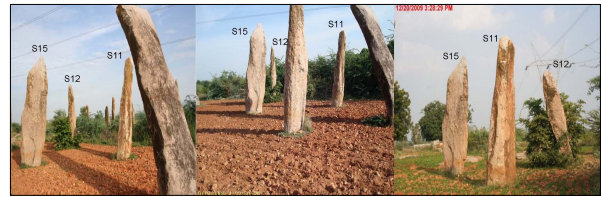

Figure 21: Evening shadows of stone s11 on s15 and s12 during (a) the summer solstice (note that the shadow of s11 is at the base of s12); (b) the equinox (the shadow of s11 is between s12 and s15); and (c) the winter solstice (the shadow of s11 is at the base of s15). The azimuths and altitudes of the Sun were calculated as 289.39° and 20.74° in (a); 264.05° and 26.3° in (b); and 230.72° and 28.67° in (c), respectively.

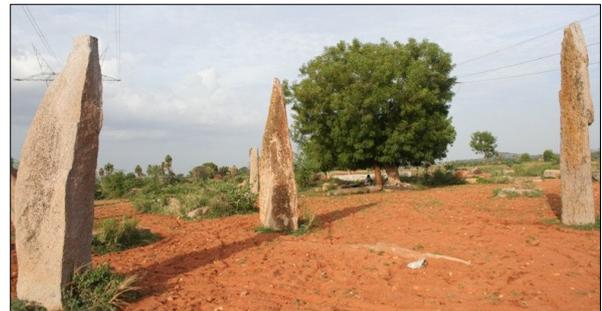

Figure 22: The shadow of stone s21 on s22 during the evening at the summer solstice. The Sun's calculated azimuth is 289.38° and altitude is 20.83°.

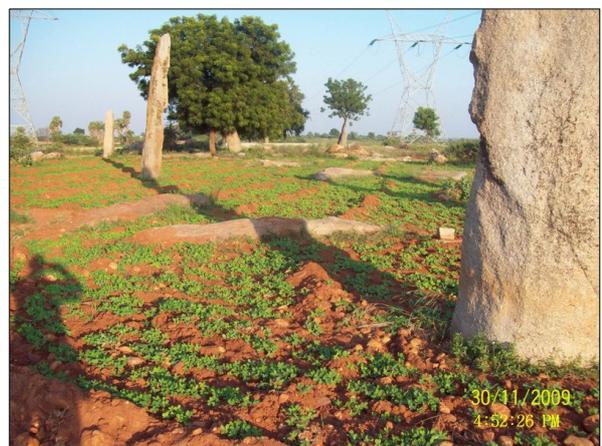

Figure 23: Stone s25 casts a shadow at the base stone s20 (slightly to the left), and s20, in turn, casts its own shadow slightly to the left of the base of s19, forming a long continuous dark path on the evening of 30 November 2009 (20 days before the winter solstice). The azimuth and altitude of Sun as calculated are 243.34° and altitude is 10.85°.

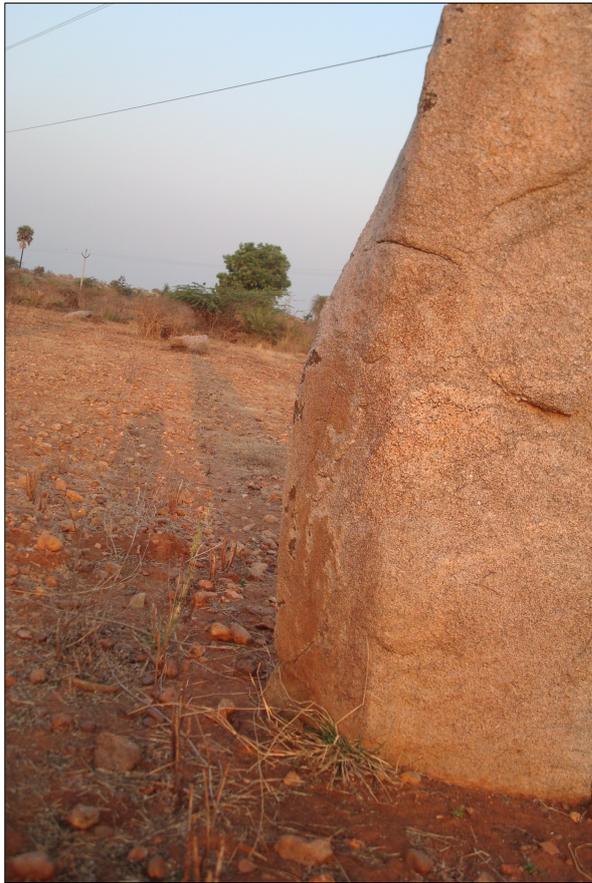

Figure 24: The shadow of the tall menhir s19 falling on a stone in one of the smaller stone arrays that surrounds Nilurallu at sunrise on the day of the equinox (21 March 2011), illustrating that the small stones are also arranged along east-west lines. Note that the extension of the north-south line defined by the tall stones also passes through the small stones.

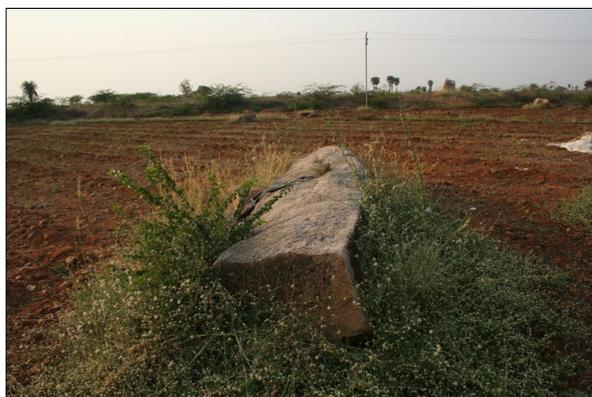

Figure 25: The base of a fallen stone is shown. Note the flat bottom and the shaping of the stone. Considerable effort might have been put into shaping the stone, probably by beating with smaller stones.

Observations of sunrises and sunsets would provide a count of the day of the year (or from the equinox or solstice), whereas the shadows of the stones would indicate the specific time of day. Unlike at Vibhuthihalli, all of the natural horizon is clearly visible from the Nilurallu stone alignment.

### 4.3 Uniqueness of the Nilurallu Stone Alignment

The Nilurallu alignment is unique and is surrounded by smaller stone arrays (particularly to the south). The spacing of the stones in the array seems to be consistent with the standard measurement of 37 ± 3 feet (11.28 meters) (see Paper I) noted at Vibhuthihalli and other Indian stone alignment sites.

The smaller stones immediately surrounding the Nilurallu site are also aligned with the large menhirs at Nilurallu that point in the directions of E-W and N-S. This is illustrated, for example, on the equinox day when the rising Sun casts the shadows of stones s19 and s17 on the small stones on the periphery of the Nilurallu complex (see Figure 24). It looks as though a grid of small stones pointing to the east and west and to the north and south was prepared before the large heavy stones were erected in the proper directions. Thus, Nilurallu became an integral part of its astronomical surroundings.

Measurements at Nilurallu must have evolved for a profound purpose from the earlier period. There do not seem to be any other stone alignment sites with such tall stones listed in Allchin's survey, although individual menhirs as long as 25 feet (or 7.6 meters) are mentioned.

Answers to the questions of where the stones come from, how they were shaped and transported, and how they were erected are not clear. Our survey of the terrain immediately surrounding the site did not reveal any stone quarries, which indicates that the stones originated from somewhere else. However, there are prominent rock outcrops in the general area (i.e. within a kilometre of the site) that could have provided stones up to 18 feet in length. Some of these were near the banks of the Krishna River. The question then arises as to how the stone slabs were extracted from the outcrop. In a few places there are indications that wooden(?) pegs may have been used to increase cracks in the rock and cause it to fracture.

As can be seen in Figure 25 (a fallen stone), the base of the stone is flat and also seems to have been shaped, maybe by pounding with hammer stones. Magli (2009: 14) mentions a method of shaping stone during the megalithic period:

> … so, the quarrying and shaping of the stones was done with tools made of stone. If the quarried stone was relatively soft, like limestone, one could easily use tools made of harder stone. However, for stones like granite or andesite (which is similar to granite, and found in the Andes), one has to use "percussors," which were chunks of the same material worked roughly into spheres and violently thrown against the area to be removed.

We found what look to be 'percussors' at Nilurallu (see Figure 26), but whether they were used as suggested by Magli is not known. A granite stone 4 meters in height and 1.2 meters in radius would weigh ~48 tons, so it is a major challenge to determine how such stones could have been transported to the megalithic site and erected. Pebbles on the ground might have offered help when dragging the big stones. The subject of moving large stones in ancient times has been discussed by Magli (2009), but the specific method that was used at Nilurallu is not clear. We did not find any evidence that iron tools were ever used.

### 4.4 The Age of the Nilurallu Stone Alignment

The site provides evidence for the existence of cultures of different eras. In Paper I we discuss various factors that suggested that the Vibhuthihalli stone alignment site dated to period between 1400 and 1800 BC. The Nilurallu site must date to a later period for various reasons mentioned previously. The requirement for more precise time measurements may have been a driving force at Nilurallu. The technology for handling large stones and erecting them in fairly accurate arrays also suggests a later date. The considerable planning, labour and devotion involved in building the monument suggest a major and very important purpose. Since there is no evidence that iron tools were used (at least not in any major way) the alignment could have been constructed sometime between 1400 and 1000 BC (before the emergence of the Iron Age in India), but of course this is only an estimate, and it is based upon very limited 'hard' evidence. The location and the dominance of the monument over its surroundings suggest a great and sacred purpose and a society that was technically more proficient than the one found earlier at Vibhuthihalli.

## 5 CONCLUDING REMARKS

The Nilurallu stone alignment is a remarkable monument, and we conclude that its primary (if not sole) purpose was to serve as a calendrical device. Sunrise and sunset observations and the patterns of the shadows of stones were used to measure time, days and fractions of a day. The regular need for a good calendar (for agricultural and other purposes) may have been the driving force behind the erection of this monument, but there could also have been ritualistic and other purposes involved as well. Considerable knowledge of engineering and astronomy was required for the successful construction of this megalithic structure.

A serious archaeological study is required to accurately determine the date of this monument, and methods like archaeomagnetism might be helpful in this regard.

Whether night time astronomy was ever pursued at Nilurallu is not clear. The claim that a figure of the Great Bear is depicted on a nearby rock cannot be taken seriously as the numerous cup marks on the rock surface show no resemblance whatsoever to the constellation that we see in the night sky.

### 5.1 Preservation of the Site

Whatever the purposes of the Nilurallu stone alignment site—and we strongly suggest that astronomy was one of them—it is a remarkable monument that demonstrates the technological advancement and the skill of the megalithic people in this region. It is a pity that such an impressive structure is slowly being destroyed: stones are being removed, and because the land is under cultivation the stability of the stones is threatened. During the two years or so that our study was carried out several stones were removed from the site, and its long-term survival is under threat. The area is at present privately owned, but we would urge relevant Government and other agencies to now make every effort to preserve this unique heritage site.

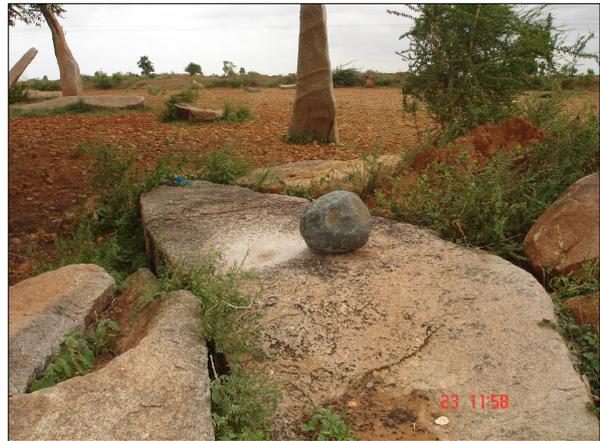

Figure 26: A small spherical stone ('percussor'?) sitting on a tall fallen stone. The spherical stone might have been used as a hammer-stone in shaping the larger stones

## 6 ACKNOWLEDGEMENTS

We would like to thank one of the anonymous referees for several helpful suggestions and comments that improved the paper. We acknowledge with thanks the help received from the Archaeological Survey of India, Bangalore, in permitting us to use their library.

Several people helped us in conducting the field work, but most notably Mr A.V. Manohar Reddy. Our sincere thanks are also due to our colleagues who are working on this project, Drs A. Vagiswari, Christina Birdie, A.V. Raveendran and A.B. Vergese, for their help and encouragement. One of the authors (NKR) would also like to thank Monique Gomez from the Instituto de Astrofisica de Canarias, Tenerife, for tracking down the abstracts of the two papers by Pulla Rao.

Finally, we wish to thank the Department of Science and Technology, Government of India, for their financial support through Project SR/S2/HEP-26/06.

N. Kameswara Rao was a Visiting Scientist at the Instituto de Astrofisica de Canaries, Tenerife, from June to Augusdt 2011, where some refinements of this paper were carried out. He retired from the Indian Institute of Astrophysics (IIA), Bangalore, as a Senior Professor of Astrophysics in 2007. His main research interests are hydrogen-deficient stars, R CrB stars, observational studies of stellar evolution and circumstellar dust, and the history of observational astronomy in India. He is also presently the PI of a Department of Science and Technology project on the development of observational astronomy in India. He is a member of the International Astronomical Union and the Astronomical Society of India.

Priya Thakur is an Assistant Professor at the Postgraduate Department of History and Archaeology at Tumkur University in Karnataka. She obtained her Ph. D from the University of Mysore in Ancient History and Archaeology. She previously worked as a Project Assistant at the IIA. Her research interests lie mainly in archaeoastronomical studies, archaeology, art history and epigraphy. She has published more than twelve research papers. Priya is associated with the Ancient Sciences and Archaeological Society of India, the Epigraphical Society of India and the Indian Art History Congress.

Yogesh Mallinathpur was a Project Assistant at the IIA and worked with Kameswara Rao. He is also doing a Ph.D at Deccan College Post Graduate Research Institute in Pune. His main research interest is in the prehistoric sites of Southern India. He is a life member of Ancient Sciences and Archaeological Society of India.